\documentclass[useASM, referee]{mn2e}
\usepackage{graphicx}

\usepackage{lscape}
\usepackage{lineno}
\usepackage{amssymb}
\usepackage{setspace}

\begin{document}
%\linenumbers
\singlespacing

\title[Precession Motion of Magnetar in GRB 180620A]
{Early Evolution of a Newborn Magnetar with Strong Precession Motion in GRB 180620A}

\author[Zou et al.]
{Le Zou$^{1}$, En-Wei Liang$^{1}$\thanks{E-mail:lew@gxu.edu.cn}
\\
$^1$Guangxi Key Laboratory for Relativistic Astrophysics, School of Physical Science and Technology, Guangxi University, Nanning 530004, China\\
}
\maketitle

\label{firstpage}
\begin{abstract}
The observed early X-ray plateau in the afterglow lightcurves of some gamma-ray bursts (GRBs) is attributed to the dipole radiations (DRs) of a newborn magnetar. A quasi-periodic oscillation (QPO) signal in the plateau would be strong evidence of the magnetar precession motion. By making a time-frequency domain analysis for the X-ray afterglow lightcurve of GRB 180620A, we find a QPO signal of $\sim650$ seconds in its early X-ray plateau. We fit the lightcurve with a magnetar precession model by adopting the Markov chain Monte Carlo algorithm. The observed lightcurve and the QPO signal are well represented with our model. The derived magnetic field strength of the magnetar is $B_{\rm p}= (1.02^{+0.59}_{-0.61})\times10^{15}$~G. It rapidly spins down with angular velocity evolving as $\Omega_{s} \propto(1+t/\tau_{\rm sd})^{-0.96}$, where $\tau_{\rm sd}=9430$~s. Its precession velocity evolution is even faster than $\Omega_s$, i.e. $\Omega_{ p}\propto (1+t/\tau_{p})^{-2.18\pm0.11}$, where $\tau_{p}=2239\pm206$~s. The inferred braking index is $n=2.04$. We argue that the extra energy loss via the magnetospheric processes results in its rapid spin-down, a low braking index of the magnetar, and the strong precession motion.
\end{abstract}
\begin{keywords}
gamma-ray burst: individual (GRB 180620A), stars: magnetars
\end{keywords}

\section{Introduction}

A rapidly spinning millisecond magnetar is a potential candidate for the central engine of gamma-ray bursts (GRBs; Usov 1992; Dai \& Lu 1998a, 1998b; Zhang \& M\'{e}sz\'{a}ros 2001; Metzger et al. 2011). Strong evidence for the magnetar engine is the detection of a steady, long-lasting X-ray plateau in the early X-ray afterglow lightcurves of about half of GRBs with the X-ray telescope (XRT) on board the $Swift$ mission (Burrows et al. 2005). This is consistent with the temporal evolution of the injected kinetic luminosity of the dipole radiations (DRs) from a spinning down magnetar, i.e. $L_{\rm k}=L_{\rm k,0}(1+t/\tau_{\rm sd})^{\nu}$, where $L_{\rm k,0}$ is the initial kinetic luminosity and $\tau_{\rm sd}$ is the characteristic spin-down timescale (e.g., Dai \& Lu 1998a, 1998b; Zhang \& M\'{e}sz\'{a}ros 2001). In the case of $\nu={-1}$, the magnetar spin-down is dominated by the gravitational wave (GW) radiation, and the inferred braking index of the magnetar ($n$) is $5$. In the case of $\nu=-2$, the spin-down is governed by the electromagnetic (EM) emission via the DRs, and the inferred braking index is $n=3$ (Ostriker \& Gunn 1969; Goldreich \& Julian 1969; Abbott et al. 2008). The EM emission arises from an internal energy dissipation process, being independent of the external shocks of the GRB fireball. This is convincingly confirmed with the observed very sharp drop with a flux decaying slope being steeper than -3 followed the X-ray plateau in some early X-ray lightcurves of GRBs, such as GRB 070110A, GRB 060607A, and 101225A (Troja et al. 2007; Liang et al. 2007;  L\"{u} \& Zhang 2014; Zou et al. 2019, 2021a). It was proposed that such an internal plateau provides evidence for collapse of a supra massive magnetar to a black hole (Troja et al. 2007; Liang et al. 2007; L\"{u} et al. 2014).

It was suggested that a newborn magnetar might experience strong precession motion in its early stage. The strong magnetospheric torque can drive the evolution of the inclination angle ($\alpha$) between the rotational axis \textbf{$\Omega$} and the magnetic axis $\textbf{B}$, leading to the precession motion of the magnetar (Stella et al. 2005; Dall'Osso \& Stella 2007; Philippov et al. 2014; \c{S}a\c{s}maz Mu\c{s} et al. 2019). If the magnetar is rapidly rotating and highly elliptical, its precession may modulate the injected kinetic luminosity of the DRs, leading to observable flux oscillation in the lightcurve (e.g. Suvorov \& Kokkotas 2020). Zou et al. (2021b) suggested that GRB 101225A is an off-axis GRB powered by a supra massive magnetar, and its early soft gamma-ray plateau is attributed to the strong DRs of the magnetar, similar to CDF-S XT2 and X-ray transient 210423 discovered with Chandra X-ray Telescope (Xue et al. 2019; Lin et al. 2021). They found QPO features in the X-ray flares of GRB 101225A at a late epoch ($t>10^3$ seconds) and proposed that the QPO features are resulted from the strong precession motion before the collapse of the magnetar to a black hole. Due to the lack of its early X-ray observations and the early soft gamma-ray being weak, they cannot reveal the information of early evolution of its precession motion of the magnetar in GRB 101225A.

GRB 180620A is of interest since its X-ray afterglow lightcurve is composed of significant flares. Globally, the lightcurve is illustrated as a plateau followed by a steep decay segment. Becerra et al. (2019) argued that the plateau is attributed to the late energy injection to the afterglow fireball. However, the peculiar flares and the steep decay of the X-ray lightcurve motivate us to suspect that the X-ray data may be a distinct emission component from the afterglows of the GRB fireball. We assume that these flares are produced from an internal dissipation process. Upon the temporal features of the X-ray lightcurve, we propose that the X-ray afterglows are attributed to the DRs of a newborn magnetar with strong precession motion, which results in the observed X-ray flares.

We dedicate to studying the early precession evolution of the magnetar in GRB 180620A in this letter. We present our temporal analysis of its X-ray data in Section 2. We fit the X-ray lightcurve with a magnetar precession model by utilizing a Markov Chain Monte Carlo (MCMC) algorithm in Section 3. Discussions and conclusions are presented in Sections 4 and 5, respectively. Throughout, a concordance cosmology with parameters $H_{0} = 70$ ${\rm km}$ ${\rm s}^{-1}$ ${\rm Mpc}^{-1}$, $\Omega_{M}=0.3$, and $\Omega_{\Lambda}=0.7$ is adopted.

\section{Data Analysis}

\begin{figure*}
\includegraphics[angle=0,scale=0.22]{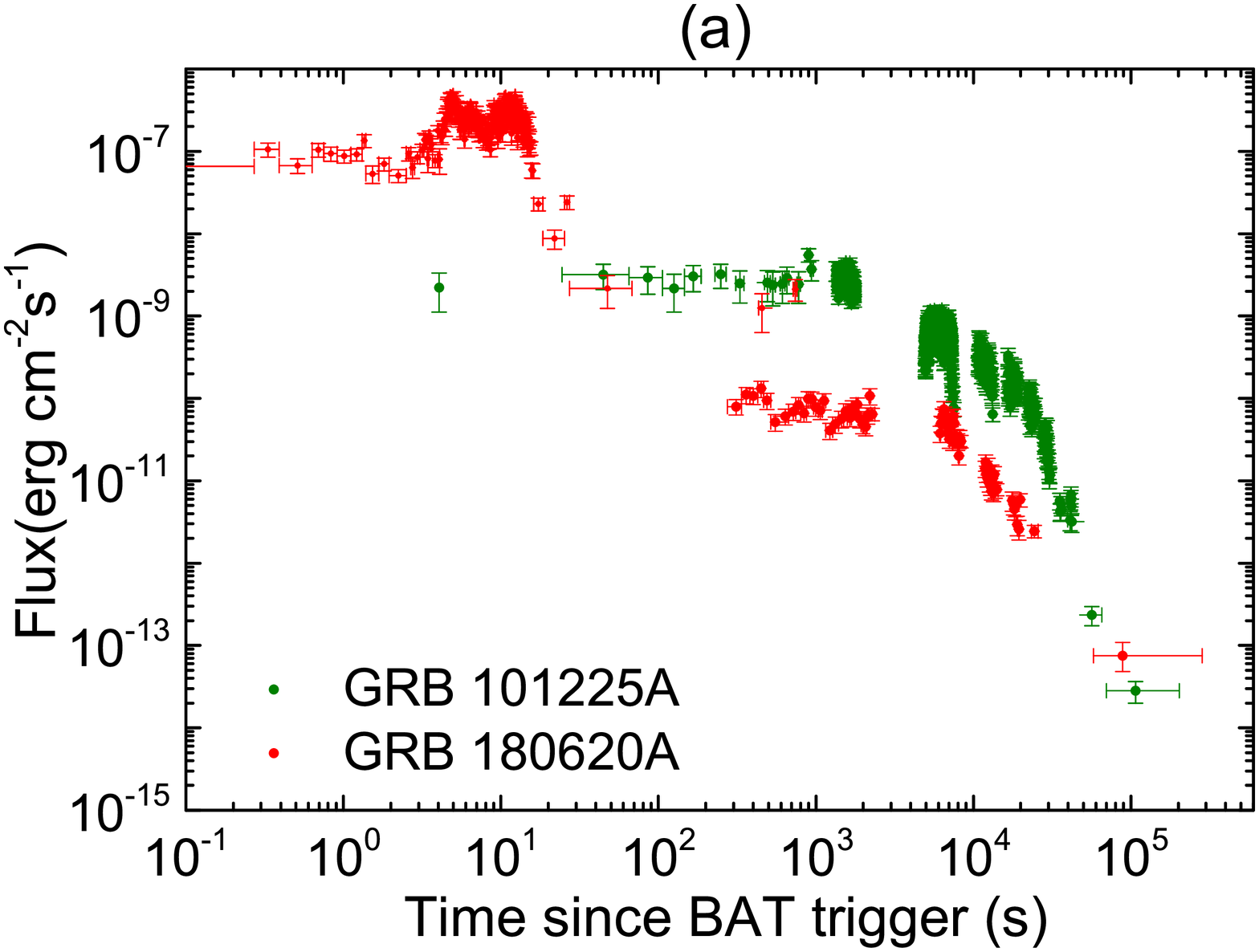}
\includegraphics[angle=0,scale=0.22]{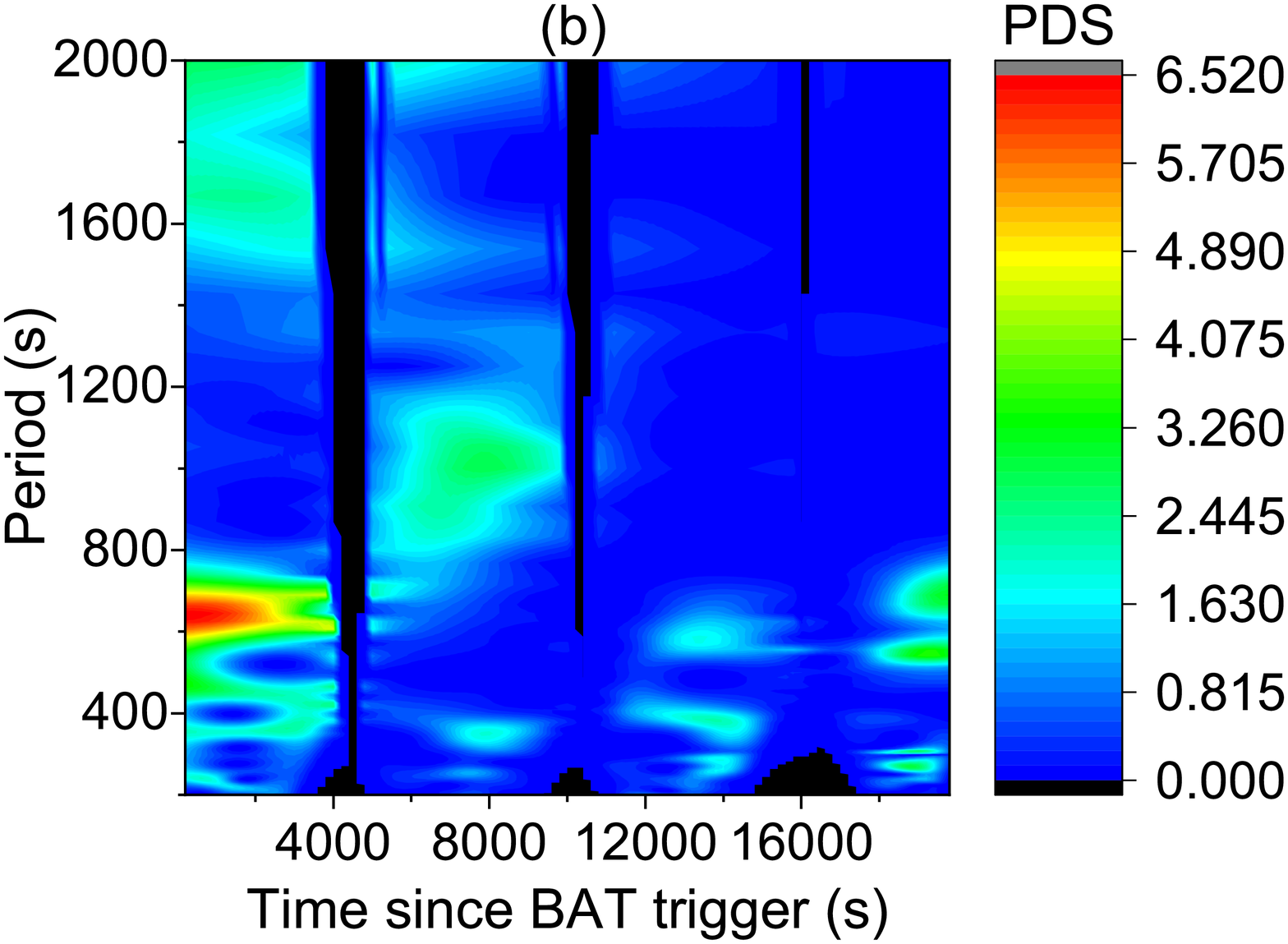}
\includegraphics[angle=0,scale=0.22]{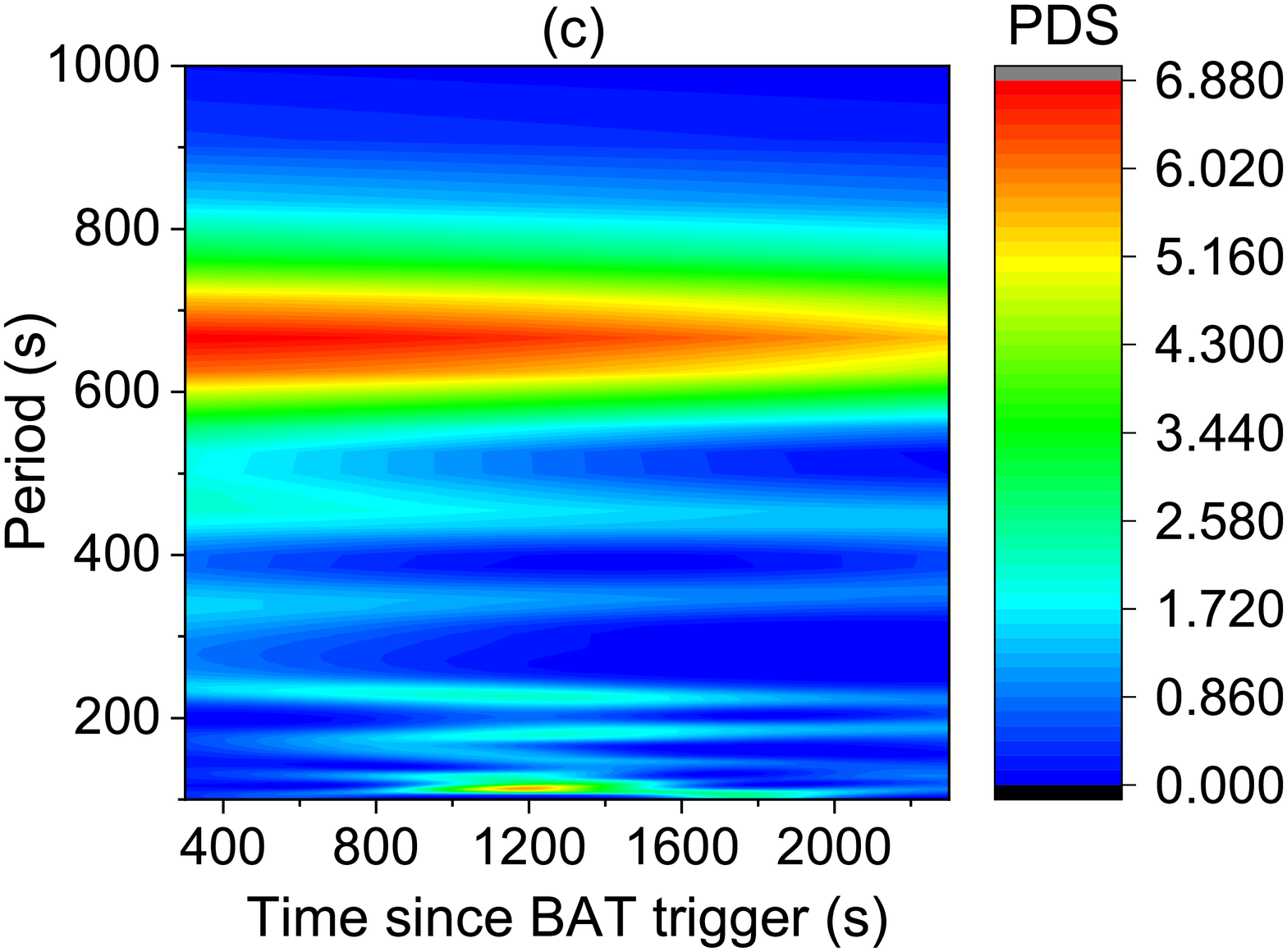}
\includegraphics[angle=0,scale=0.22]{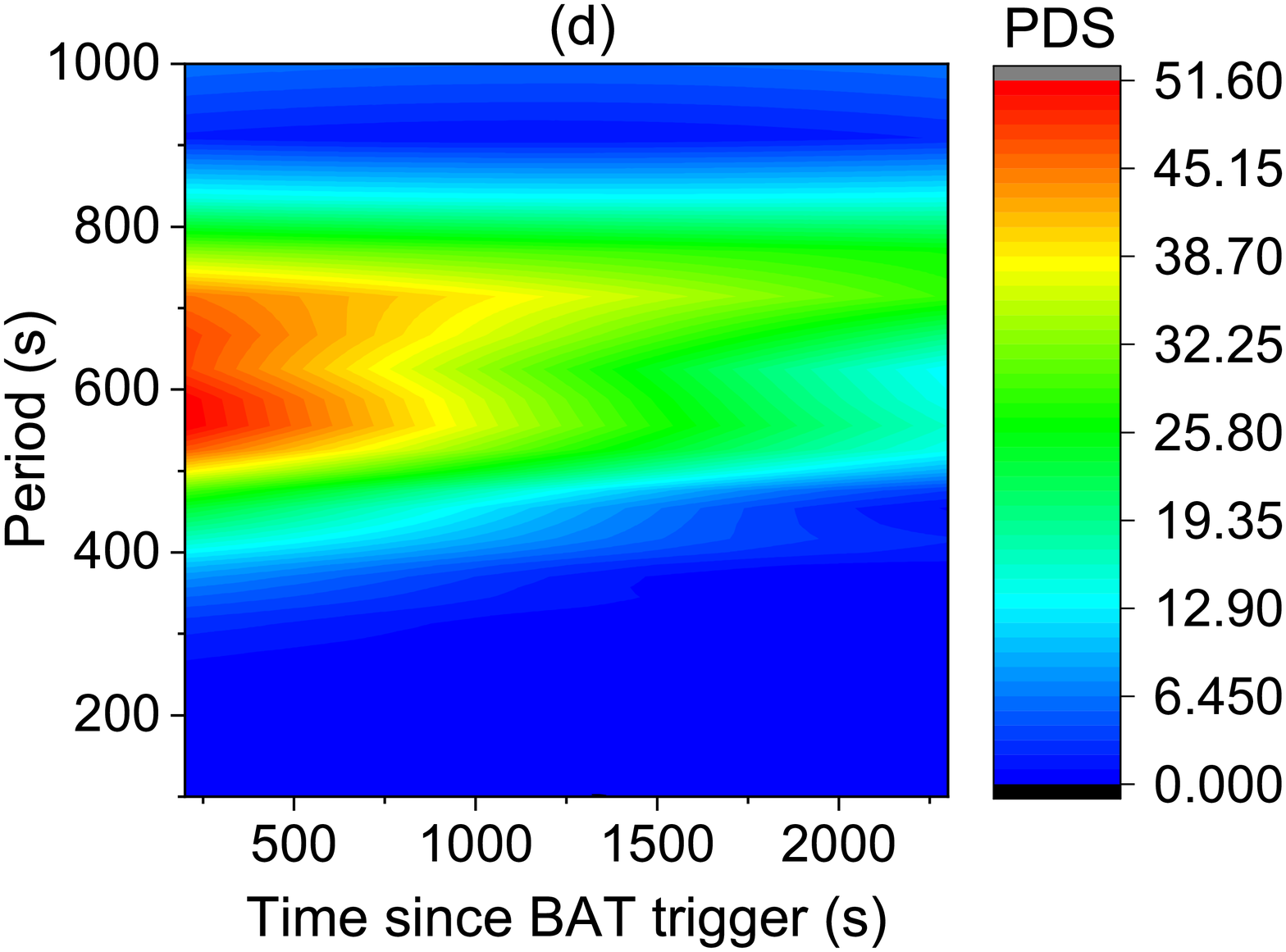}
\center\caption{Panel (a)---Prompt X-ray lightcurve extrapolated from the BAT data and X-ray afterglow lightcurves of GRB 180620A (red dots) in comparison with the gamma-ray/X-ray lightcurves of GRB 101225A (olive dots). Panels (b, c, d)---Power-density spectrum derived from our time-frequency domain analysis with the weighted wavelet Z-transform algorithm for the X-ray lightcurve of GRB 180620A: (b) global X-ray lightcurve, (c) zooming in the range of $t\in(200, 2300)$ seconds, (d) X-ray lightcurve derived from our model in the range of $t\in(200, 2300)$ seconds.
}
\label{1}
\end{figure*}

GRB 180620A was triggered by the Burst Alert Telescope (BAT) on board the $Swift$ mission on 2018 June 20 at 08:34:58 (UT dates). Its redshift limit is 1.2 (Breeveld et al. 2018). We adopt the BAT and XRT data available at the website of $Swift$ Burst Analyzer \footnote{https://www.swift.ac.uk/xrt\_curves/00843122/}
(Evans et al. 2010). The prompt and the X-ray afterglow lightcurves of GRB 180620A are shown in Fig. \ref{1}(a). Its prompt gamma-ray lightcurve shows several overlapping peaks with a total duration of about 25~s in the 15-150 keV band (Stamatikos et al. 2018). The highly variable prompt gamma-ray lightcurve indicates that GRB 180620A is an on-axis GRB. Interestingly, an extended emission component with a steady flux up to $\sim 700$ seconds is also observed in the BAT band.

The X-ray afterglow lightcurve is composed of flares, even at its late epoch up to $10^{4}$ seconds. Globally, the lightcurve is initially featured as a plateau then transits to a steep decay segment at $t\sim 10^{4}$ seconds without considering the flux variations of the flares. The photon index of the time-integrated spectrum of the plateau is $\Gamma_X=1.34^{+0.07}_{-0.06}$ by fitting the spectrum with an absorbed power-law spectrum. The X-ray spectrum is harder than that usually seen in the X-ray afterglows of the GRB fireball (e.g., Liang et al. 2007). Our time-resolved spectral analysis for each X-ray flare in the plateau does not find a spectral evolution signature. The X-ray lightcurve of GRB 180620A is similar to the gamma-ray/X-ray lightcurves of GRB 101225A, which was suggested to be powered by the DRs of a supra massive magnetar (Zou et al. 2021b). We also show the gamma-ray/X-ray lightcurves of GRB 101225A in Fig. \ref{1} (a) in comparison with GRB 180620A. The detection of bright, highly variable prompt gamma-ray indicates that GRB 180620A is an on-axis GRB, and the lack of detection of such a gamma-ray emission component likely implies that GRB 101225A is an off-axis GRB. A soft gamma-ray plateau (extended gamma-ray emission) accompanied by the X-ray plateau is observed in these two GRBs.

We search for possible QPO features of the flares in the plateau of GRB 180620A. Considering that the angular velocity ($\Omega_p$) of the magnetar precession should evolve with time, we adopt the weighted wavelet Z-transform (WWZ) (Foster 1996; Gong et al. 2022) to make time-frequency domain analysis for searching possible QPO signatures in the lightcurve. The analysis result is shown in Fig. Fig. \ref{1}(b). A QPO signal of $P=650\pm 50$ seconds is found at $t\in(200,2300)$ seconds. We zoom in the QPO signal in Fig. \ref{1}(c). We estimate the confidence level of the QPO as described in Timmer and Koenig (1995) and find that the QPO signal has a confidence level of 3$\sigma$.

\section{Magnetar Precession Model}

The global temporal feature and the QPO signature of the X-ray data are likely to result from the DRs of a  magnetar with strong precession motion (Suvorov \& Kokkotas 2020; Zou et al. 2021b). In this scenario, the observed X-ray lightcurve is given by
\begin{equation} \label{eq:EM}
L_X(t)= \eta_X L_{\rm k,0}(1+\frac{t}{\tau_{\rm sd}})^{\nu}\lambda,
\end{equation}
where $\eta_X$ is the radiation efficiency of the ejecta in the XRT band (0.3-10 keV), $\lambda$ is a magnetospheric factor, and $\nu$ is defined as $\nu\equiv 4/(1-n)$. In the EM dominated scenario, we have
\begin{equation}
L_{\rm k,0,49} = 1.0 {B^2_{\rm p,15}P^{-4}_{\rm s,0,-3}R^{6}_{6}},
\end{equation}
\begin{equation}
\tau_{\rm sd,3}=2.05 I_{45}B^{-2}_{\rm p,15}P^{2}_{\rm s,0,-3}R^{-6}_{6},
\end{equation}
where $B_{\rm p}$ is the dipole magnetic field strength, $P_{\rm s, 0}$ is the initial spin period, $R$ is the radius, $I$ is the inertial moment of the magnetar, and $Q_m=Q/10^m$ is in the cgs units.

The magnetospheric factor $\lambda$ depends on the orientation of the NS (Goldreich \& Julian 1969; Spitkovsky 2006; Kalapotharakos \& Contopoulos 2009; Philippov et al. 2015). We adopt a hybrid model $\lambda(\alpha)\simeq1+\delta \sin^{2}\alpha$, where the parameter $\delta$ quantifies the magnetospheric physics ($|\delta|\leq 1$) (Philippov et al. 2014; Arzamasskiy et al. 2015; Suvorov \& Kokkotas 2020). The $\alpha$ evolution is $\dot{\alpha}\approx k \Omega_{p}\csc\alpha\sin(\Omega_{p} t)$.
The $\lambda(\alpha)$ evolves over time as (Goldreich 1970; Zanazzi \& Lai 2015; Zou et al. 2021b)
\begin{equation} \label{eq:alpha}
\lambda(\delta,\alpha_0,k,\Omega_p)=1+\delta \sin^{2}\alpha \approx 1 + \delta[1-(\cos\alpha_{0} + k( \cos(\Omega_{p} t)-1))^{2}],
\end{equation}
where $\alpha_{0}$ is the initial inclination angle, $k$ is an order-unity factor related to the other Euler angles, $\Omega_p$ is the precession rotational frequency. We attribute the observed QPO signature to the precession motion. Estimating the $\Omega_p$ value with the QPO, we get $\Omega_p=0.010$ rad/s. Since $\Omega_p$ would evolve with time, we parameterized the evolution of $\Omega_p$ as $\Omega_{p}(t) \propto \Omega_{p,0} (1+t/\tau_{p})^{\varphi}$, where $\tau_p$ is a characteristic timescale of the $\Omega_p$ evolution.

\begin{figure*}
\includegraphics[angle=0,scale=0.25]{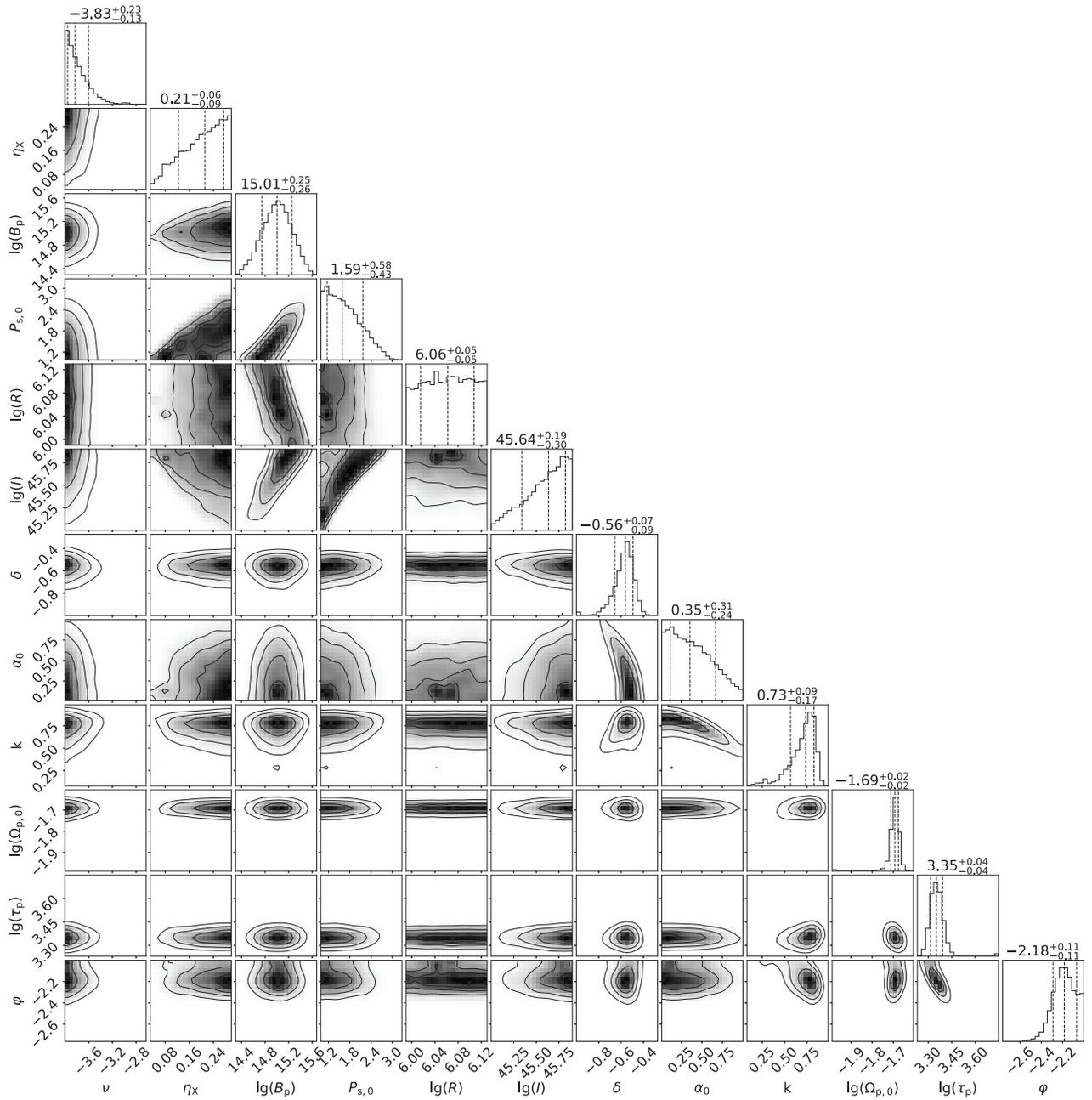}
\center\caption{ Probability contours of the model parameters derived from our MCMC fit for GRB 180620A. The vertical dashed lines mark the $1\sigma$ confidence level regions centering at their median probabilities.
}
\label{2}
\end{figure*}

\begin{figure*}
\includegraphics[angle=0,scale=0.25]{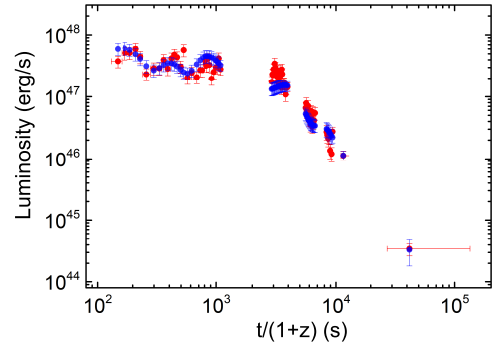}
\center\caption{X-ray lightcurve (blue dots) derived from our best model fit in comparison with the observed X-ray afterglow lightcurve of GRB 180620A (red dots). The model lightcurve is sampled the same as the observations and their uncertainties are also taken the same as the observed ones.}
\label{3}
\end{figure*}

We calculate the X-ray lightcurve in the burst frame, as shown in Fig. \ref{3}, where the isotropic X-ray luminosity is calculated as $L_{\rm iso,X}=4\pi D^{2}_{L}F(t^{'})(1+z)^{\Gamma{_X}-2}$ and $t^{'}=t/(1+z)$. We fit the lightcurve based on Eqs. (1)-(4). The free parameters of our model are $\nu$, $\eta_X$, $B_{\rm p}$, $P_{\rm s,0}$, $R$, $I$, $\delta$, $\alpha_0$, $k$, $\Omega_{p, 0}$, $\tau_{p}$, and $\varphi$. We adopt the \texttt{emcee} python package (Foreman \& Mackey et al. 2013) based on the MCMC algorithm to fit the data for constraining these parameters. The derived probability distributions of our fits are shown in Fig. \ref{2}, where the vertical dashed lines mark the region of $68.3\%$ ($1\sigma$) centering at the median probability (the middle vertical dashed line). In case of the probability distribution is a Gaussian-type function, the parameter is well constrained. Otherwise, the parameters are poorly constrained or only a limit can be obtained. It is found that $B_{\rm p}$, $\delta$, $k$, $\Omega_{p,0}$, $\tau_p$, and $\varphi$ are well constrained, i.e. $B_{\rm p}= (1.02^{+0.59}_{-0.61})\times10^{15}$~G, $\delta=-0.56^{+0.07}_{-0.09}$, $k=0.73^{+0.09}_{-0.17}$, $\Omega_{p,0}=0.02\pm 0.01$~rad/s, $\tau_{p}=2239\pm206$~s, and $\varphi=-2.18\pm0.11$, where the uncertainties are estimated in $1\sigma$. The parameters $\nu$, $\eta_X$, $P_{\rm s,0}$, $R$, $I$, and $\alpha_0$ cannot be constrained. We take their values as that at the median probability, i.e. $\nu=-3.83$, $\eta_X=0.21$, $P_{\rm s,0}=1.59$~ms, $\alpha_{0}=0.35$~rad, $R=1.15\times10^{6}$ $\rm cm$ and $I=4.37\times10^{45}$ $\rm g$ $\rm cm^{2}$.
We show our fitting curve in Fig. \ref{3} by sampling the model lightcurve and adopting the uncertainty the same as the observed data. One can observe that the data can be represented with our model.

Note that the conversion efficiency of the spin-down luminosity to the X-ray luminosity is an important parameter in constraining the $B_{\rm p}$ and $P_{\rm s,0}$. Our analysis shows $B_{\rm p}$ is well constrained, but $P_{\rm s,0}$ and $\eta_X$ are poorly constrained. Based on Eqs. (2) and (3), we have $\tau_{\rm sd, 3}L_{\rm k,0,49}=2.05I_{45}P^{-2}_{\rm s, 0, -3}$. Thus, $\eta_X$ can be estimated with $\eta_X=0.49 \tau_{\rm sd, 3}L_{X,0,49}I^{-1}_{45}P^{2}_{\rm s, 0, -3}$. The $\tau_{\rm sd}$ value is theoretically determined by $B_{\rm p}$ and $P_{\rm s,0}$, but the break time of the X-ray plateau ($t^{'}_{b}$) in the burst frame can be roughly taken as a lower limit of $\tau_{\rm sd}$. In addition, the break-up spin period is also a robust lower limit of $P_{\rm s,0}$, i.e. $P_{\rm s,0}>0.98$ ms (Lattimer \& Prakash 2004) and adopting $I=4.37\times10^{45}$ $\rm g$ $\rm cm^{2}$. Thus, we can set a lower limit of $\eta_{\rm X}>0.03$ with the initial X-ray luminosity $L_{X,0}=6\times 10^{47}$ $\rm erg/s$ and $t^{'}_b\sim 4000$ seconds. The derived $\eta_X$ in our MCMC fit does not violate this limit. It is comparable to that reported by Du et al. (2016). A high radiation efficiency implies that the saturated Lorentz factor of the ejecta powered by the DRs should be low (Xiao \& Dai 2019).

\section{Discussion}

\subsection{Low Braking Index and Rapid Spin-Down of the Magnetar}

Our fit gives $\nu=-3.83^{+0.23}_{-0.13}$, which indicates a braking index of $n=2.04^{+0.07}_{-0.03}$. We know that $n={5}$ for the spin-down of the magnetar purely dominated by the GW emission and $n=3$ for the EM emission dominated scenario. The inferred $n$ value is smaller than that in the EM dominated scenario. Note that the spin velocity evolves as $\Omega_{s}(t)=\Omega_{ s,0}(1+t/\tau_{\rm sd})^{1/(1-n)}$. We get $\Omega_s(t)\propto (1+t/\tau_{\rm sd})^{-0.96}$ by taking $n=2.04$, indicating that the decrease of $\Omega_{s}$ is much faster than that of the EM or GW dominated scenario.

Lyne et al. (2013) and Lyne et al. (2015) proposed that the evolution of the inclination angle $\alpha$ could cause $n<3$. It is possible that the low braking index is related to the precession motion. The evolution of magnetosphere torques may change the inclination angle and lead to the magnetar precession motion. Several models were proposed to explain the influence of the magnetosphere on the braking index. It was suggested that twisted magnetospheres, which consist of a solid mixed poloidal-toroidal field, may increase the spin-down torque and lead to reduce the braking index (Thompson et al. 2002; Kiuchi et al. 2011; Turolla et al. 2015). Contopoulos \& Spitkovsky (2006) proposed that the spin-down torque is also enhanced by a dipole force-/twist-free magnetosphere. In addition, the braking index should be also reduced if the co-rotating magnetosphere is small (Contopoulos \& Spitkovsky 2006) or the magnetosphere conductivity increases with time (Li et al. 2012). If the magnetar braking torque is mainly affected by its magnetosphere and the magnetosphere does not co-rotate with the central magnetar, the magnetospheric processes could exert extra torque on the crust, resulting in a loss of rotational energy of the magnetar. Such extra energy lost via the magnetospheric processes may result in its rapid spin-down and a low baking index of the magnetar in GRB 180620A.

\subsection{Precession Evolution of the Magnetar}

The precession evolution of the magnetar is governed by the magnetospheric factor $\lambda(\delta,\alpha_0,k,\Omega_{p,0})$. Our best fit yields $\delta=-0.56^{+0.07}_{-0.09}$, $\alpha_0=0.35^{+0.31}_{-0.24}$~rad, $k=0.73^{+0.09}_{-0.17}$, and $\Omega_{p} =10^{-1.69\pm 0.02} (1+t/\tau_{p})^{-2.18\pm 0.11}$~rad/s, where $\lg \tau_p=3.35\pm {0.04}$. $\Omega_{p}$ evolves from $2.00\times 10^{-2}$ at $t=0$ to $4.51\times 10^{-3}$ at $t=\tau_{p}$, corresponding to a period from 308 seconds to 1394 seconds. A QPO signal can be observed in the early stage ($t<\tau_p$) since $\Omega_p$ does not significantly evolve in the early stage. Our QPO analysis reveals a QPO signal of $P=650\pm 50$ seconds in the time slice $t\in(200,2300)$ seconds. The $\Omega_p$ value derived from our model at $t=900$ second, roughly the middle of the slice $t\in(200,2300)$ seconds, is $635$ seconds, being consistent with the observed one. We also make a QPO analysis for the lightcurve derived from our model in the same time interval. Our result shows in Fig. \ref{1}(d) for comparison with that for the data. The QPO signal of $P=600\sim 750$ seconds is found. It agrees with the observed one, although the signal spreads in a broader time range than the observed one. This would be due to the temporal evolution of $\Omega_p$. Since $\Omega_p$ rapidly decreases at the late time epoch ($t>\tau_p$), no QPO signal should be found.

Note that the $\Omega_{p}$ evolution depends on the ellipticity ($\epsilon$) and the spin angular velocity ($\Omega_{s}$) as $\Omega_{p}(t) \approx \epsilon (t) \Omega_{s}(t)$ (Jaranowski et al. 1998). Thus, the initial $\epsilon_{0}$ can be obtained as $\epsilon_{0} \approx 5.17\times 10^{-6}$.  We also obtain $\tau_{\rm sd}=9430$~s from Eq. (3). It found that $\tau_{\rm sd}$ is longer than  $\tau_{ p}$ by a factor of 4.21, indicating that the decreases of $\Omega_{p}$ is faster than $\Omega_{s}$. Note that $\epsilon$ depends on the deformation of a newborn magnetar, which is mainly induced by its magnetic field and starquake (Mastrano et al. 2018; Giliberti \& Cambiotti 2022). The change of the magnetic field can be ignored in the early stage since its decay timescale is relatively long, and the deformation induced by the magnetic field can be regarded as a constant in the early stage (Turolla et al. 2015). The starquake-induced deformation is related to its spin. Therefore, the evolution of $\epsilon$ should be the same as $\Omega_{s}$.

\section{Conclusions}

We have revisited the X-ray afterglow lightcurve of GRB 180620A regarding significant flares in the early X-ray plateau and steep decay post the plateau. Our analysis in the time-frequency domain for its X-ray lightcurve reveals a QPO signature of $P\sim650$ seconds with a $3\sigma$ confidence level in the time interval of $t\in(200,2300)$ seconds. We explain these features with the DRs of a newborn magnetar with a strong precession, in which the EM radiation dominates the spin-down. We show that our model well represents the global temporal feature and the QPO signal. By constraining the model parameters with the MCMC algorithm, we find that $B_{\rm p}= (1.02^{+0.59}_{-0.61})\times10^{15}$~G and the median value $P_{\rm s,0}=1.59$~ms of the magnetar. It spins down with angular velocity evolving as $\Omega_s \propto(1+t/\tau_{\rm sd})^{-0.96}$, which is much faster than the pure EM dominated scenario. The inferred braking index is 2.04. We argue that extra energy lost via the magnetospheric processes results in its rapid spin-down, a low baking index of the magnetar, and the strong precession motion. Its the median value of initial inclination angle is $\alpha_{0}=0.35$~rad and the evolution of the precession angular velocity is $\Omega_{p}=0.02(1+t/\tau_{p})^{-2.18\pm0.11}$~rad/s, where $\tau_{p}=2239\pm206$~s. Being due to the rapid evolution of $\Omega_p$, no QPO signal can be detected at $t>\tau_p$, which is consistent with the observations.

One caveat should be addressed that a plateau-like segment in the well-sampled optical afterglow lightcurves of GRB 180620A was also observed, as shown in Figure 1 of  Becerra et al. (2019), who explained the plateau in both the optical and X-ray bands as a late energy injection to the external shock of the GRB fireball. The W band optical lightcurve has the best temporal coverage (from $t=40$ seconds to $t=9500$ seconds) for our discussion. It is composed of a smooth bump (rising slope $-0.83\pm0.07$ and decaying slope $1.32\pm0.01$) in the slice $t\in (100,1000)$ seconds, another broad bump (rising slope $-0.23\pm0.08$ and decaying slope $0.36\pm0.02$) in $t\in (1000\sim 6200)$ seconds, and a power-law decay segment with a slope of $-1.01\pm0.24$ at $t>6200$ seconds. These temporal features indeed can be explained with the reverse and forward external shock models as proposed by Becerra et al. (2019). The broad optical bump may also be explained as a spectral break of synchrotron radiation, as shown in Figure 2 of Sari \& Piran (1999). These features are distinct from the flares observed in the X-ray lightcurve. In addition, the optical data observed in $t=8\times 10^3$ seconds in the i and r bands indicate that the lightcurves should have a break in the time interval from $t=8\times 10^3$ to $t=8\times 10^4$ seconds. Becerra et al. (2019) assumed the break at $t=8\times 10^3$ seconds and obtained an optical decay slope of $\alpha_{\rm O}=-1.84\pm0.12$ in this time interval. However, the lack of observations in this time interval makes the slope quite uncertain since it highly depends on the break time. In the X-ray band, Becerra et al. (2019) reported a flux decay slope of $\alpha_X\sim -2.17\pm 0.13$ at $t>8\times 10^{3}$ seconds. Note that the source is still tentatively detected with a flux of $\sim 7\times 10^{-14}$ at $t\sim 10^5$ second. The X-ray flux decay slope by considering this data point is steeper than $-2.17$.
We obtain a much steeper slope ($-3.83^{+0.23}_{-0.13}$) from our model fit by considering the XRT observation at $t>8\times 10^{4}$ seconds. Such a steep decay slope implies that the X-ray emission should be from an internal energy dissipation process. Thus, whether both the optical and X-ray afterglows have the same physical origin or not is ambiguous.

\section*{Acknowledgements}

We appreciate thoughtful comments and suggestions from the referee. We also thank Can-Min Deng, Yun-Lu Gong, and Ji-Gui Cheng for useful discussion. We acknowledge the use of the public data from the {\em Swift} data archive and the UK {\em Swift} Science Data Center. This work is supported by the National Natural Science Foundation of China (Grant No.12133003 and U1731239), Guangxi Science Foundation (grant No. 2017AD22006), and Innovation Project of Guangxi Graduate Education (YCBZ2020025).

%**********************************************************************

\section*{Data availability}

The data underlying this article are available in the article.

\clearpage

\end{document}